\documentclass[fleqn,12pt,twoside]{article}
\usepackage[headings]{espcrc1}

\readRCS $Id: espcrc1.tex,v 1.2 2004/02/24 11:22:11 spepping Exp
$ \usepackage{graphicx}
\usepackage{epsfig}
\usepackage[figuresright]{rotating}

\newcommand{\AmS}{{\protect\the\textfont2  A\kern-.1667em\lower.5ex\hbox{M}\kern-.125emS}}

\def\b0{{\mbox{\boldmath$0$}}}

     \font\tenbifull=cmmib10 scaled 1200 
     \font\tenbimed=cmmib9
     \font\tenbismall=cmmib7
       \def\bmit{\fam9 }
\mathchardef\bbkappa="7114
\mathchardef\bbrho="711A
\mathchardef\bbsigma="711B
\mathchardef\bbtau="711C
\mathchardef\bbvarrho="7125
\mathchardef\bbvarsigma="7126
\mathchardef\bbxi="7118

\def\boldrho{{\bmit\bbrho}}

\def\b0{{\mbox{\boldmath$0$}}}

\def \b #1{ {\bf #1}}
\newcommand{\be}{\begin{eqnarray}}
\newcommand{\ee}{\end{eqnarray}}

\def \b #1{ {\bf #1}}

\newcommand{\CM}{{\cal M}}

\def \b #1{ {\bf #1}}

\title{Hadron Propagation in  Medium: the Exclusive Process A(e,e'p)B in Few-Nucleon Systems
       }

\author{C. Ciofi degli Atti\address[a]{Department of Physics, University of Perugia and Istituto
        Nazionale di Fisica Nucleare, Sezione di Perugia, Via A. Pascoli, I-06123, Italy},
        L. P. Kaptari\addressmark[a]\thanks{On leave from Bogoliubov Lab. Theor. Phys., 141980, JINR, Dubna, Russia}
        and
        H. Morita\addressmark[a]\thanks{On leave from Faculty of Social Information, Sapporo Gakuin University,
        11 Bunkyo-dai, Ebetsu-shi, Hokkaido 069-8555, Japan}}

\runtitle{Hadron Propagation in the Medium} \runauthor{H. Morita}

\begin{document}

\maketitle

\begin{abstract}
The mechanism of propagation of hadronic states in the medium is a
key  point for understanding particle-nucleus and nucleus-nucleus
scattering at high energies. We have investigated the propagation of
a baryon in the exclusive process A(e,e'p)B in few-nucleon systems
using realistic nuclear wave functions  and Glauber multiple
scattering theory both in its original form and within a generalized
eikonal approximation. New results for the processes
$^3He(e,e'p)^2H$ and $^4He(e,e'p)^3H$ are compared with data
recently obtained at the Thomas Jefferson Laboratory (JLAB).
\end{abstract}

\section{Introduction}
Exclusive and  semi-inclusive lepton scattering off nuclei $A({\it l},{\it
l}'p)X$ in the quasi elastic region,  plays a relevant role in nowadays
hadronic physics,  mainly for three reasons: i) due to the wide  kinematical
range available by present experimental facilities  non trivial information on
nuclei (e.g. nucleon-nucleon (NN) correlations) can be obtained; ii) the
mechanism of propagation of hadronic states can be investigated in great
details; iii) at high energies color transparency effects might be
investigated. The key point here is a reliable evaluation of the mechanisms of propagation of
the produced
hadrons in the medium, a task which is usually referred
to as the  problem of the evaluation of  the Final State
Interaction (FSI).  At medium and high energies
 hadron propagation is usually treated within the  Glauber
multiple scattering approach (GA) ,  which has been applied with great success
to  hadron scattering off nuclear targets \cite{glauber}. However, when the
hadron is created inside the nucleus,  as in a process $A({\it l},{\it l}'p)X$,
various improvements of the original GA have been advocated. Most of them are
based upon a Feynman diagram reformulation of GA; such a diagrammatic approach,
 has been developed long ago
for the case of hadron-nucleus scattering \cite{gribov} and it  has been
generalized to the process $A({\it l},{\it l}'p)X$ \cite{FrankStrik97,Misak05},
showing that in particular kinematical  regions it predicts appreciable
deviation from GA. In such an approach, based upon a generalized eikonal
 approximation  (GEA) the   frozen approximation, common to
GA, is partly removed by taking
 into account the excitation energy of the  $A-1$ system,
  which results in a correction term to the standard
 profile function of GA,  leading   to an additional contribution to the
  longitudinal component of the missing momentum.
The GEA has recently  been applied  to a systematic calculation of the exclusive
processes $^3He(e,e'p)^2H$ and $^3He(e,e'p)(np)$ \cite{CiofiRev,CiofiLet} using
realistic three-body  wave functions \cite{Kiev} and  two-nucleon
interactions (AV18) \cite{AV18}; the results of calculations  show a nice agreement with
recent Thomas Jefferson Laboratory (JLAB) data \cite{E89044}. The two-body
break up channel  $^3He(e,e'p)^2H$ has also been considered within
the Glauber approach in \cite{rocco},
obtaining results consistent with Ref. \cite{CiofiRev,CiofiLet}. The  aim of
this contribution is twofold: i) to extend the GEA calculation to the four-body
system, namely to the calculation of  the process $^4He(e,e'p)^3H$, for which recent data
have been obtained at JLAB \cite{E97111}; ii) to consider for the same reaction, through the
concept of Finite Formation Time (FFT) as developed in Ref. \cite{Braun}, the
role played by nucleon virtuality   which is expected to become
important at high values of $Q^2$. Our  paper is organized as follows: in
Section 2 the basic elements  of our
theoretical framework are  given; the comparison of our results with
experimental data on  $^3He(e,e'p)^2H(pn)$ and $^4He(e,e'p)^3H$ reactions are
presented  in Section 3; FFT effects on  the process $^4He(e,e'p)^3H$ are
illustrated in Section 4; the  Summary and Conclusions  are  given in Section 5.

\section{The cross section for the process  A(e,e'p)B  within GA and GEA}

The FSI which is considered in the  diagrammatic  approach of
Ref. \cite{FrankStrik97,Misak05,CiofiRev,CiofiLet}
is the elastic scattering of the hit nucleon by the nucleons of the
spectator $A-1$. Under two main assumptions which
are expected to be valid at medium and high energies, namely that: i)
in each
rescattering process the momentum transfer is small,  and ii) the spin
flip part of the NN scattering amplitude
can be disregarded, the method predicts that nuclear effects in the
exclusive process $A(e,e'p)B$ should be governed by the Distorted Spectral Function
\begin{eqnarray}
P_A^{FSI}({{\bf p}_m},E_m)&=& \frac {1}{(2\pi)^3} \frac{1}{2J_A+1}\,\,\,
\sum_{f}\,\,\, \sum_{{\mathcal M}_A,\,{\mathcal M}_{A-1},\,\,s_1} \left |
\sum_{n=0}^{A-1} {\mathcal T}_A^{(n)}({\mathcal M_A},{\mathcal M}_{A-1}, s_1;f)
\right |^2\times \nonumber\\
&\times& \delta\left( E_{m}-(E_{A-1}^f + E_{min})\right)
\label{pdistor}
\end{eqnarray}
where ${\bf p}_m={\bf P_{A-1}} = {\b q} - {{\b p}_1}$ and $E_m = E_{min} +  E_{A-1}^f$
are the {\it missing momentum} and
 {\it missing energy}, respectively (here ${{\b p}_1}$ and ${\b q}$ are the momentum
 of the detected nucleon and
  the 3-momentum transfer, respectively,
 and $E_{min}= E_{A}- E_{A-1}$, $E_A$ and  $E_{A-1}$ being the positive
  ground state energies of $A$ and $A-1$);  ${\cal M}_A$,  ${\cal M}_{A-1}$,
  and $s_1$,
  are magnetic quantum numbers;
 the sum over $f$ stands for a sum over  all possible discrete and
 continuum states of the
 $A-1$ system; ${\mathcal T}_A^{(n)}$ represents the the reduced
  (Lorentz index independent) amplitude
 which, at order $n$, takes into account all possible diagrams describing
 $n$-body rescattering (see \cite{CiofiRev}).
 After the evaluation of all single and double
 scattering diagrams,  the distorted spectral function  of  $^3He$ reads as follows

 \be
P_3^{FSI}({{\bf p}_m},E_m)&\hspace{-0.5cm}=&\hspace{-0.5cm}\frac{1}{2(2 \pi)^3}
  \hspace{-0.2cm}\sum_{f=D,np}\sum_{\CM_3,\CM_2,s_1}
  \left | \int {\rm e}^{i{\boldrho}\cdot{\b p}_m}
 \chi_{\frac12 s_1}^{\dagger} \Psi_{f}^{{\CM_2}\,\dagger}(\b{r} )
  {\cal S}_{GEA}(\boldrho,{\bf r})
 \Psi_{He}^{\CM_3}(\boldrho,\b{r})   d\boldrho d {\bf r} \right |^2\times\nonumber\\
 &\times& \delta\left( E_{m}-(E_{A-1}^f + E_{min})\right)
\label{pfin}
\ee
\noindent where $E_{A-1}^f + E_{min}=E_{min}$,  for the two-body break up (2buu)
channel ($f=D$),  and $E_{A-1}^f + E_{min}=
{\b {t}^2}/{M_N} + E_3$,  for the three-body break up (3buu) ($f=(np))$ channel
(here ${\bf t}$ is the relative momentum of the interacting $(n-p)$ pair
in the continuum. The quantity
  ${\mathcal S}_{GEA}$ introduces  FSI and has the form
 $ {\mathcal S}_{GEA}=
{\mathcal S}_{GEA}^{(1)}+{\mathcal S}_{GEA}^{(2)}$, with

\begin{eqnarray}
{\mathcal S}_{GEA}^{(1)}({\boldrho},{\bf {r}})=1-\sum\limits_{i=2}^3
\theta(z_i-z_1){\rm e}^{i\Delta_z (z_i-z_1)} \Gamma (\b {b}_1-\b{b}_i)
\label{essesingle}
\end{eqnarray}

\noindent and
 \begin{eqnarray}
{\cal S}_{GEA}^{(2)}({\boldrho},\bf {r})&=&
\left[\theta(z_2-z_1)\theta(z_3-z_2){\rm e}^{-i\Delta_3(z_2-z_1)} {\rm
e}^{-i(\Delta_3+\Delta_z)(z_3+z_1)}+   \right. \nonumber \\&+&
     \left.
\theta(z_3-z_1)\theta(z_2-z_3){\rm e}^{-i\Delta_2(z_3-z_1)} {\rm
e}^{-i(\Delta_2+\Delta_z)(z_2-z_1)}\right]
 \Gamma(\b {b}_1-\b{b}_2)\Gamma(\b {b}_1-\b{b}_3)
  \label{essedouble}
 \end{eqnarray}

\noindent where  $\Delta_i=(q_0/|{\b q}|)(E_{{\b k}_i} - E_{{\b k}_i^{'}})$ and
$\Delta_z = (q_0/|{\b q}|) E_m$, ${{\b k}_i}$,  ${\b k}_i^{'}$, being nucleon
momenta before and after the rescattering.
It can be seen that  $\Delta_z$ takes into account Fermi motion and therefore partly
remove the frozen approximation. Note that when  $\Delta_i=\Delta_z=0$, the usual
GA is recovered.

\section{Calculations of the processes $^3He(e,e'p)^2H(pn)$ and $^4He(e,e'p)^3H$ Reaction}

Within the diagrammatic approach,  the differential cross section assumes a factorized
form, namely
\begin{equation}
\frac{d^6\sigma}{d\nu d\Omega_{e} dpd \Omega_p} = {\mathcal
K}\sigma_{ep}P_A^{FSI}(\mathbf{p}_m,E_m), \label{eq:crosssection}
\end{equation}
where ${\mathcal K}$ is a kinematical factor, $\sigma_{ep}$ the
electron-nucleon cross section and $\nu$ the energy transfer. We
have calculated the cross sections of the processes
$^3He(e,e'p)^2H$, $^3He(e,e'p)(np)$, and $^4He(e,e'p)^3H$ using the
well known parametrization of the profile function \be \Gamma({\bf
b}) =\frac{\sigma_{NN}^{tot}(1-i\alpha_{NN})}{4\pi b_0^2}\, e^{-{\bf
b}^2/2b_0^2} \label{profile} \ee \noindent  with all parameters
taken from Ref. \cite{PDG}. For the electron-nucleon cross section
$\sigma_{ep}$ we used the De Forest  $\sigma_{ep}^{cc1}({\bar
Q}^2,{\bf p}_m)$ cross section \cite{forest}. All two-, three-, and
four-body wave functions are direct solutions of the  non
relativistic Schr\"odinger equation, therefore our calculations  are
fully parameter free.

\begin{figure}[htb]
\begin{minipage}[htb]{77mm}
\vskip -0.5cm  \epsfysize=9.0cm\epsfbox{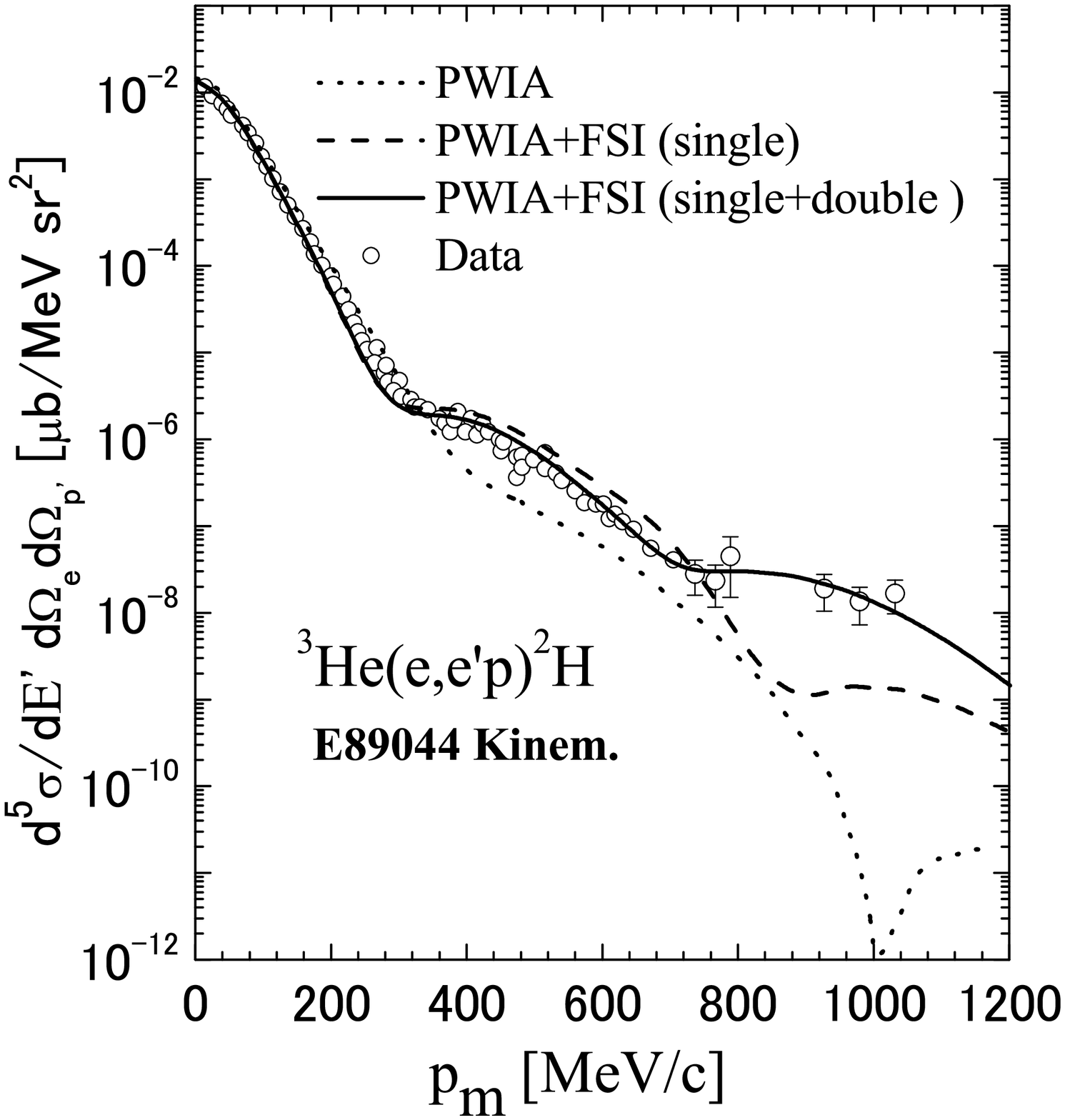}
 \vskip -1.0cm \caption{Results for the $^3He(e,e'p)^2H$ reaction \cite{CiofiLet}.
Dotted curve: PWIA result; dashed
curve:  FSI (single rescattering);
solid curve: FSI (single plus double rescattering). Experimental data from
\cite{E89044}.} \label{fig:3He2bb}
\end{minipage}
\hspace{\fill}
\begin{minipage}[htb]{77mm}
\vskip -1.0cm
      \hskip -0.8cm \epsfysize=10.5cm\epsfbox{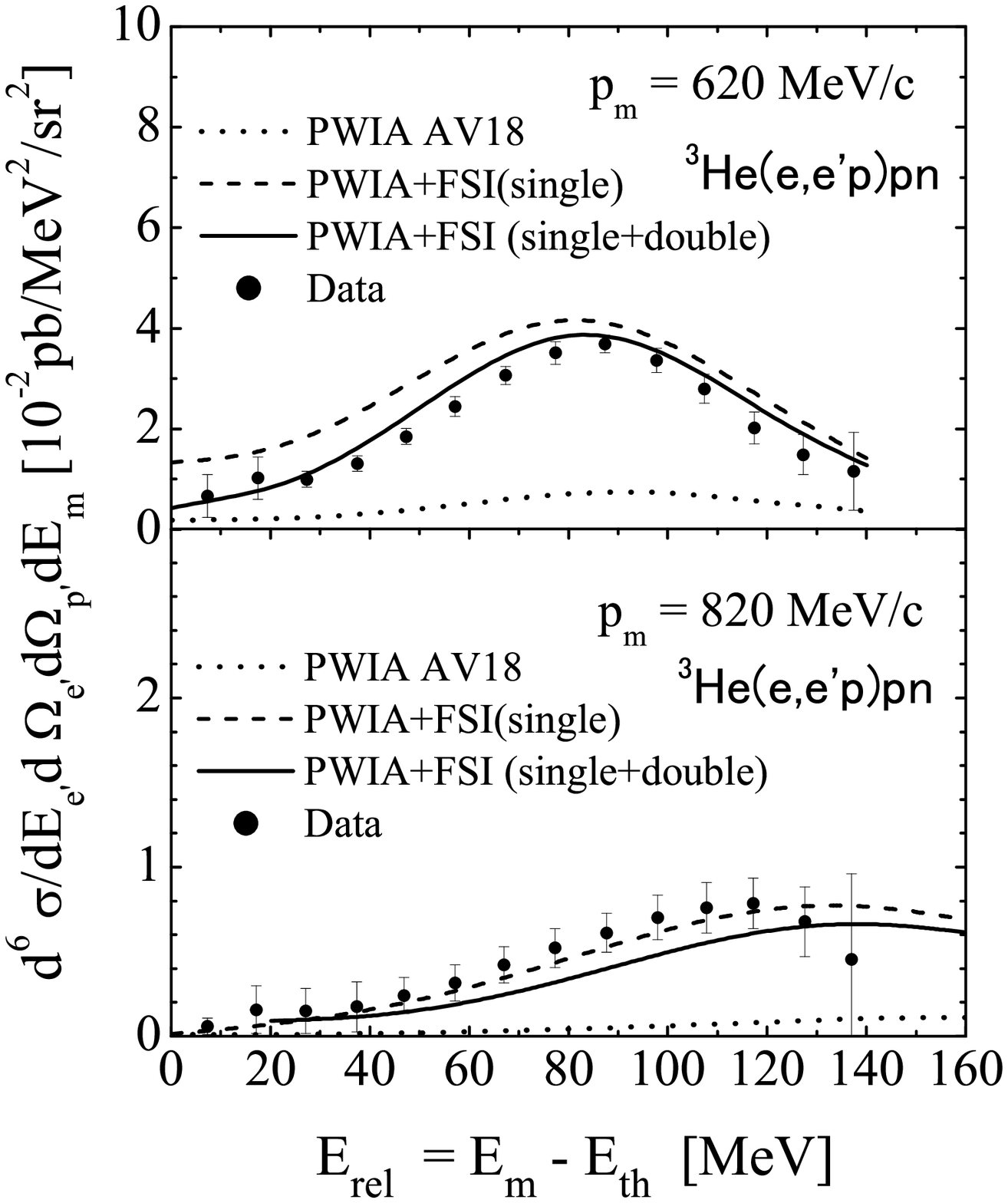}
\vskip -1.0cm \caption{The same as in Fig. \ref{fig:3He2bb} but for the process
$^3He(e,e'p)pn$ ($E_{th} =E_3$ is the two-nucleon emission threshold in $^3He$).} \label{fig:3He3bb}
\end{minipage}
\end{figure}

In case of the three-nucleon system,
the results for the 2bbu and 3bbu channels are shown in Figs. \ref{fig:3He2bb}
and \ref{fig:3He3bb} \cite{CiofiLet}. The missing momentum
\begin{figure}[htb]
\begin{minipage}[tb]{77mm} \vskip -0.0cm  \epsfysize=9.0cm\epsfbox{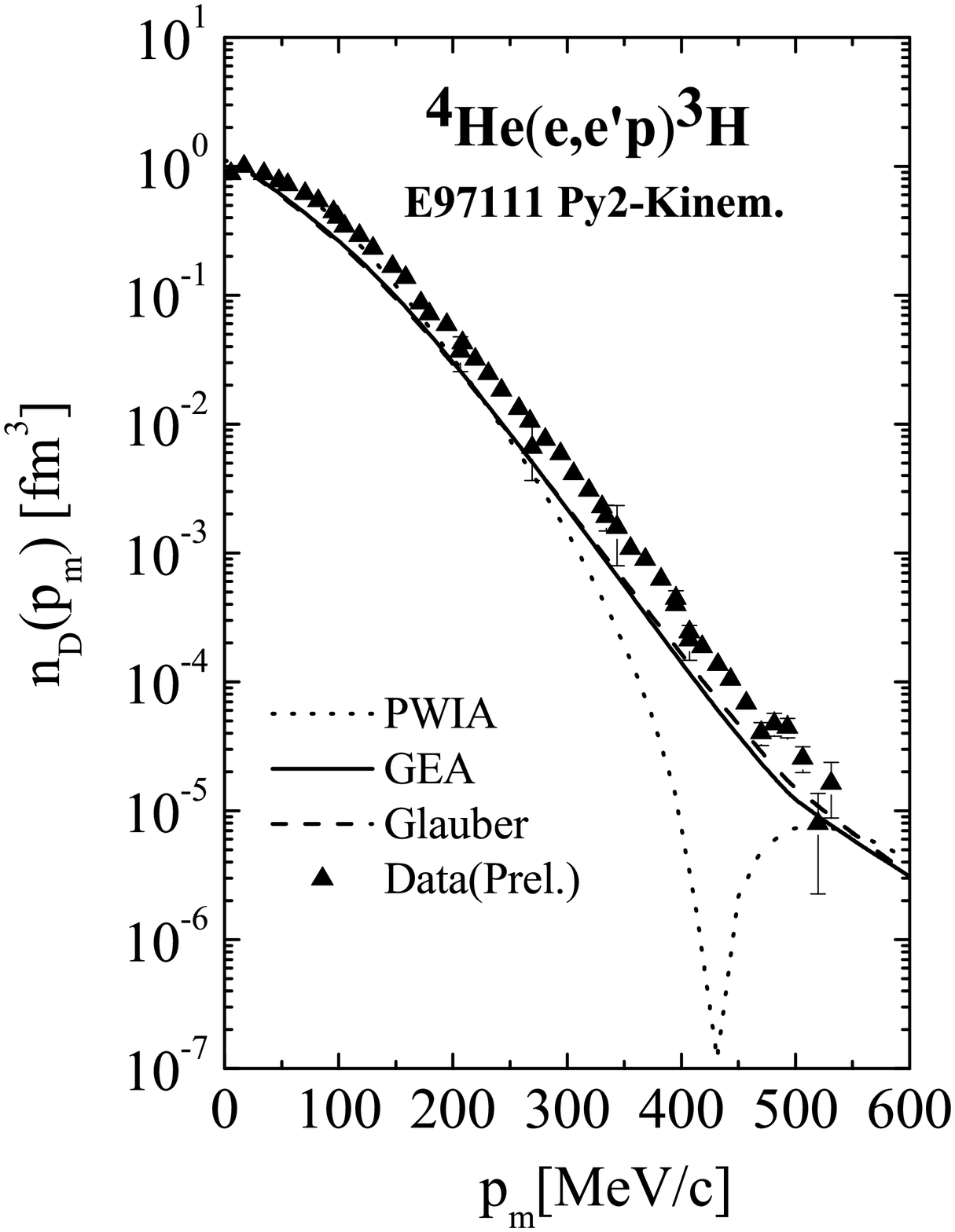}
 \vskip -1.0cm \caption{The reduced cross section
 $n_D(\mathbf{p}_m)=[d^5\sigma/(d\nu d\Omega_{e}d \Omega_p)]\times
[{\mathcal K}\sigma_{ep}]^{-1}$  for the process  $^4He(e,e'p)^3H$ in
parallel kinematics. Preliminary data from \cite{E97111}.} \label{fig:4HePy2}
\end{minipage}
\hspace{\fill}
\begin{minipage}[htb]{77mm}
\vskip -1.0cm
      \hskip -0.0cm \epsfysize=9.0cm\epsfbox{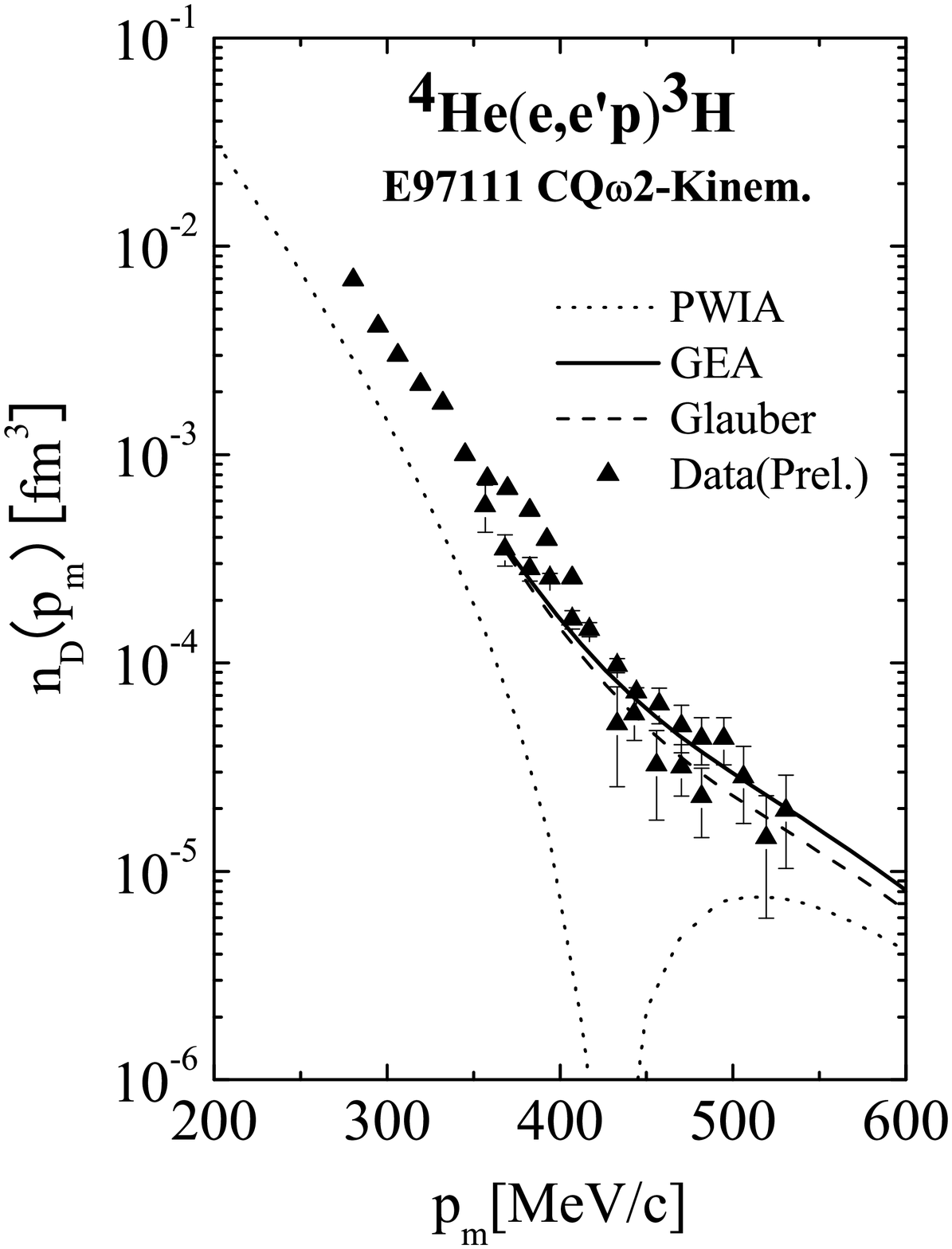}
\vskip -1.0cm \caption{The same as in  Fig. \ref{fig:4HePy2} but for
perpendicular kinematics. } \label{fig:4Hecqw2}
\end{minipage}
\end{figure}
dependence of the experimental cross section clearly exhibits
 different slopes, that are reminiscent of the slopes  observed in elastic
 hadron-nucleus scattering at
 intermediate energies (see e.g. Ref. \cite{glauber}) and
our parameter free calculations  demonstrate that: i)
 these slopes are indeed  related to multiple scattering in the final state, and
 ii)
 a highly satisfactory agreement between theory and experiment
is obtained, which means that in the energy-momentum range covered by the data,
  FSI can be described by elastic rescattering; iii)
  GA  and  GEA,
 differ only by a few percent.

  The results for $^4He$,  for which  $ {\mathcal S}_{GEA}=
{\mathcal S}_{GEA}^{(1)}+{\mathcal S}_{GEA}^{(2)}+{\mathcal S}_{GEA}^{(3)}$, are
presented for the first time in Figs. \ref{fig:4HePy2} and \ref{fig:4Hecqw2}.
In this case,
we have used realistic variational
 wave functions for both  $^4He$ and $^3H$  \cite{ATMS,ATMS2}, corresponding to the
 RSC V8 model potential\cite{RSCV8}. Calculations for the reduced cross section
\begin{equation}
n_D(\mathbf{p}_m)=\frac{d^5\sigma}{d\omega d\Omega_{e}d \Omega_p}
({\mathcal K} \sigma_{ep})^{-1},
\end{equation}
are compared with the JLab E97111 experimental data in
parallel(Py2) and perpendicular
(CQ$\omega$2)  kinematics \cite{E97111}.
\begin{figure}[htb]
\begin{minipage}[htb]{77mm}
\vskip -0.3cm  \epsfysize=9.0cm\epsfbox{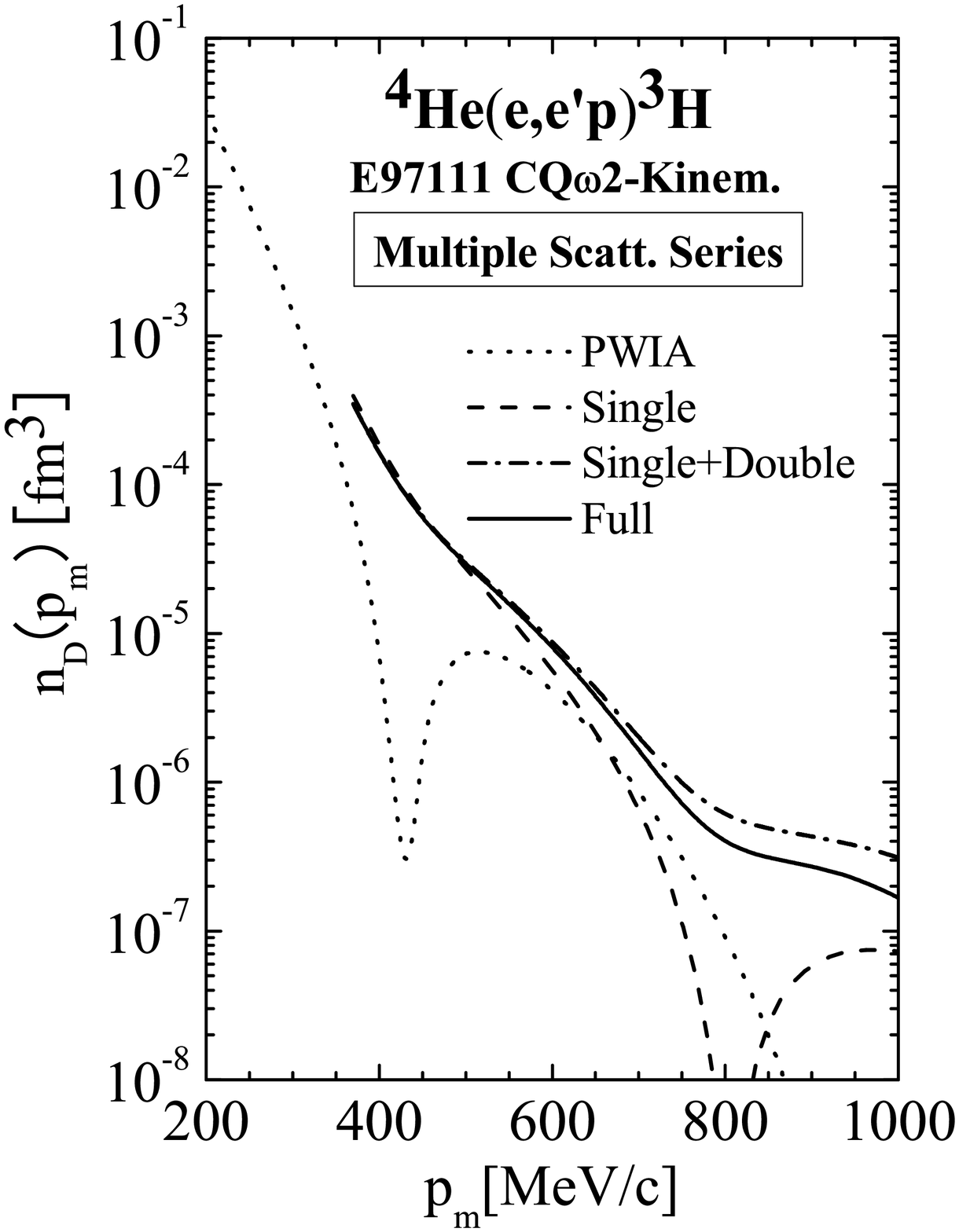}
 \vskip -1.1cm \caption{Multiple scattering contributions in the process $^4He(e,e'p)^3H$.
 The results are similar to the ones shown in Fig. \ref{fig:3He2bb},
 but in this case triple rescattering contributions start to contribute at $p_m \geq800\,MeV/c$.}
  \label{fig:4HeExp}
\end{minipage}
\hspace{\fill}
\begin{minipage}[htb]{77mm}
\vskip -0.2cm
      \hskip -0.0cm \epsfysize=9.0cm\epsfbox{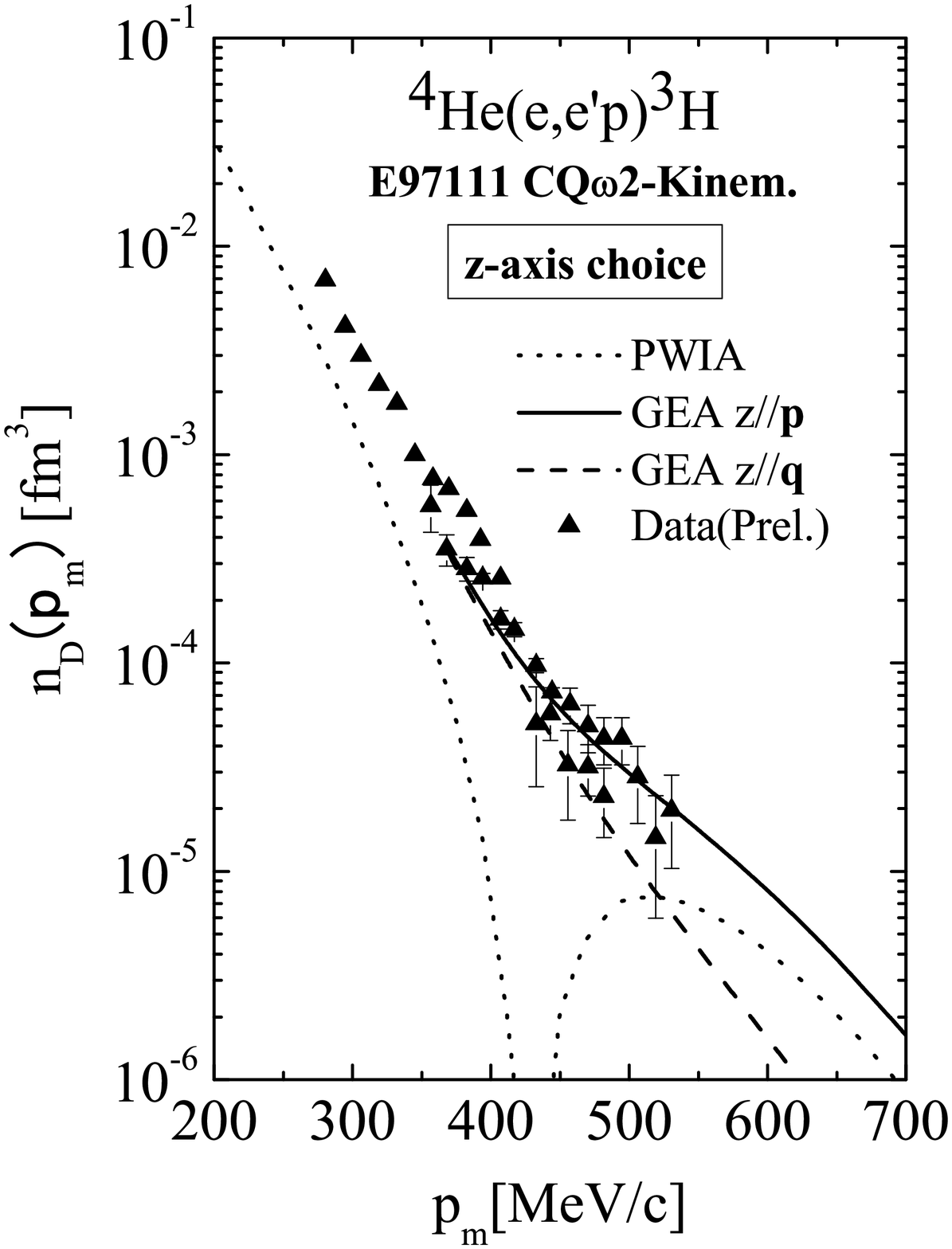}
\vskip -1.2cm \caption{The choice of z-axis for the Glauber
operator. The dashed curve corresponds to  the calculation with the
$z$-axis along the direction of \textbf{q}.  See  text. Preliminary
data from \cite{E97111}} \label{fig:4HeZaxis}
\end{minipage}
\end{figure}
Fig. \ref{fig:4HePy2} shows that: i) the dip predicted by the PWIA
 is totaly filled up  by the FSI; ii)  like the $^3He$ case, the
difference between GA and  GEA  is very small; iii)  although we predict an overall
satisfactory behaviour of the experimental data in parallel kinematics, we systematically
underestimate them.
In case of perpendicular kinematics, shown in Fig.
\ref{fig:4Hecqw2},  the agreement between theory and experiment is much better
 and the differences between GA and  GEA are more pronounced.
The multiple scattering contributions are illustrated in Fig.
 \ref{fig:4HeExp}. As the case of $^3He$ the
single rescattering amplitude dominates at $p_m\leq600\,MeV/c$ whereas at higher
values of $p_m$ multiple scattering effects  become important,  with
the triple
rescattering term  contributing significantly  at $p_m>800\, MeV/c$.
In our calculations we have always directed the $z$-axis along the momentum
of the propagating nucleon ${\bf p}_1$. In several Glauber-type
calculations the $z$-axis is chosen along  $q\mathbf{}$,
assuming $|{\bf q}|$ to be large enough. Fig.
\ref{fig:4HeZaxis} shows that this is not the case in the JLAB
kinematics, with the  calculation with  the $z$-axis directed along $\mathbf{q}$
underestimating the correct results by a large factor.

\section{Finite Formation Time Effects in the process  $^4He(e,e'p)^3H$}
It has been argued by various authors that at high values of  $Q^2$ the
phenomenon of color transparency, i.e. a reduced NN cross section in the medium,
 might be observed. Color transparency is a consequence of the cancelation
 between various hadronic intermediate states of the produced ejectile.
 In \cite{Braun} the vanishing of FSI at $Q^2$ has been produced by considering the
 finite formation time (FFT) the ejectile needs to reach its asymptotic form of a physical baryon.
 This has been implemented by explicitly considering the virtuality dependence
  of the NN scattering amplitude.
According to \cite{Braun}  FFT effects can be introduced in Eq.
(\ref{pfin}) by replacing ${\mathcal S}_{GEA}$ with ${\mathcal
S}_{FFT}$,  which is obtained from
 ${\mathcal S}_{GEA}$ simply by letting $\Delta_i=\Delta_z=0$ and replacing
 $\theta(z_i-z_1)$ by
 \be
 \textit{J}(z_i-z_1)= \theta(z_i-z_1)\left( 1-\exp [-(z_i-z_1)/{\it l}(Q^2)] \right)
 \label{eq:FFT}
 \ee
 with ${\it l}(Q^2)={Q^2}/(x m_N\,M^2)$
where $x$ is the  Bjorken scaling  variable and the quantity $l(Q^2)$ plays
the role of the proton
 formation length, the length of the trajectory that the
knocked out proton runs until it return to its asymptotic form. The quantity $M$
is related to the nucleon mass$m_N$ and to an average  resonance state of mass $m^*$
by $M^2 = {m^*}^2 - m_N^2$;
the value $m^*$ = 1.8 GeV has been used in the calculations\cite{Braun}.
Since this formation length grows linearly with $Q^2$, at higher
$Q^2$ the strength of the Glauber-type FSI is reduced by the damping
factor $( 1-\exp[-(z_i-z_1)/{\it l}(Q^2)])$ appearing  in Eq.
(\ref{eq:FFT}), which physically describes the following situation: once the
hit  proton virtually reaches  a resonance state, it will
need a finite amount of time to return to its asymptotic form,
during  which  FSI becomes weaker than the Glauber one; if $l(Q^2)$ = $0$,
 then   $S_{FFT}$  reduces
to the usual Glauber operator $S_{G}$.
\begin{figure}[htb]
\begin{minipage}[htb]{77mm}
\vskip -0.0cm  \epsfysize=9.0cm\epsfbox{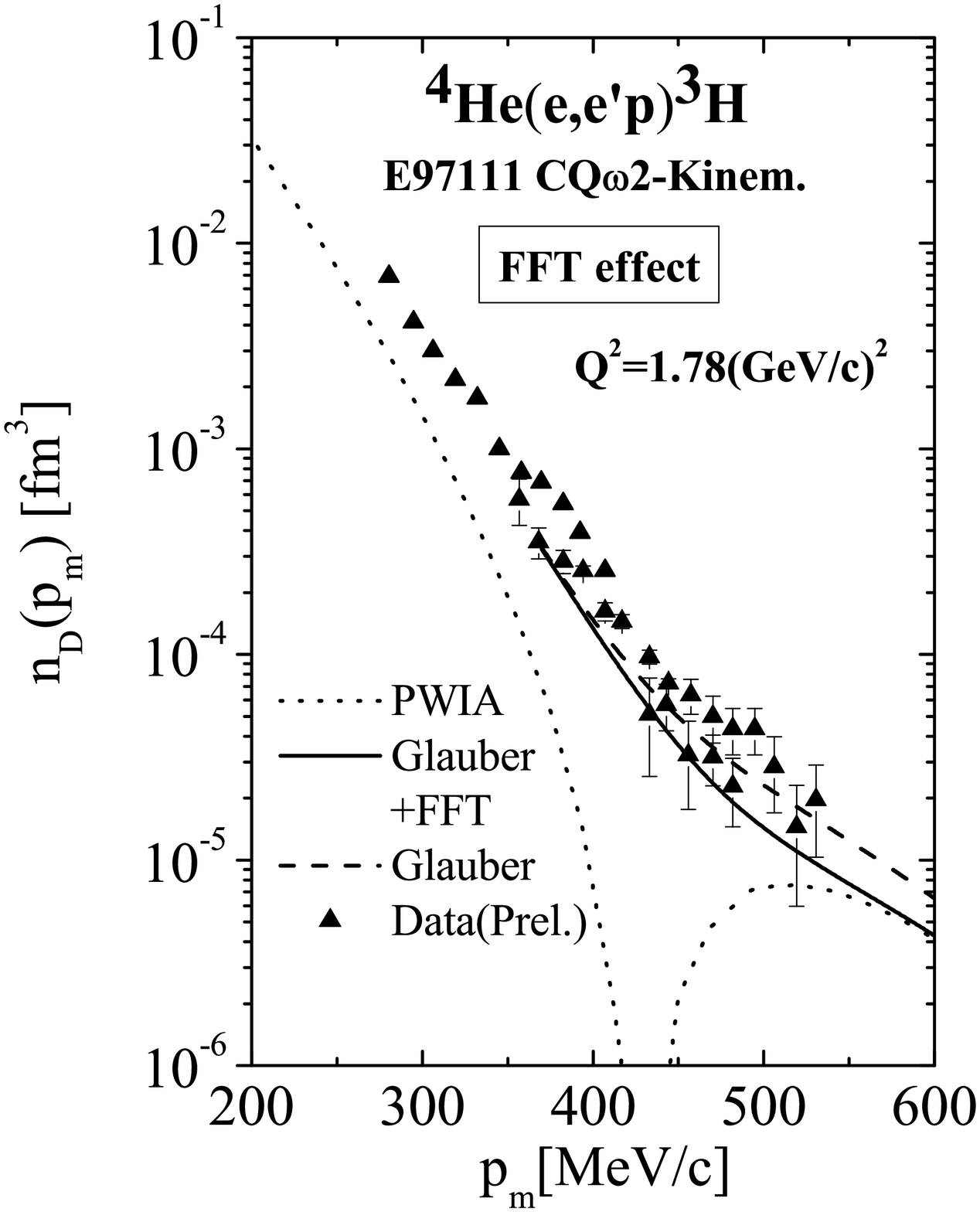}
 \vskip -1.0cm \caption{The FFT effect on the CQ$\omega2$
 kinematics. The solid line shows the results within GEA, whereas
 the dashed curve corresponds to the conventional GA.
  Preliminary  data from \cite{E97111}.}
  \label{fig:4HeFFT}
\end{minipage}
\hspace{\fill}
\begin{minipage}[htb]{77mm}
\vskip -1.0cm
      \hskip -0.0cm \epsfysize=9.0cm\epsfbox{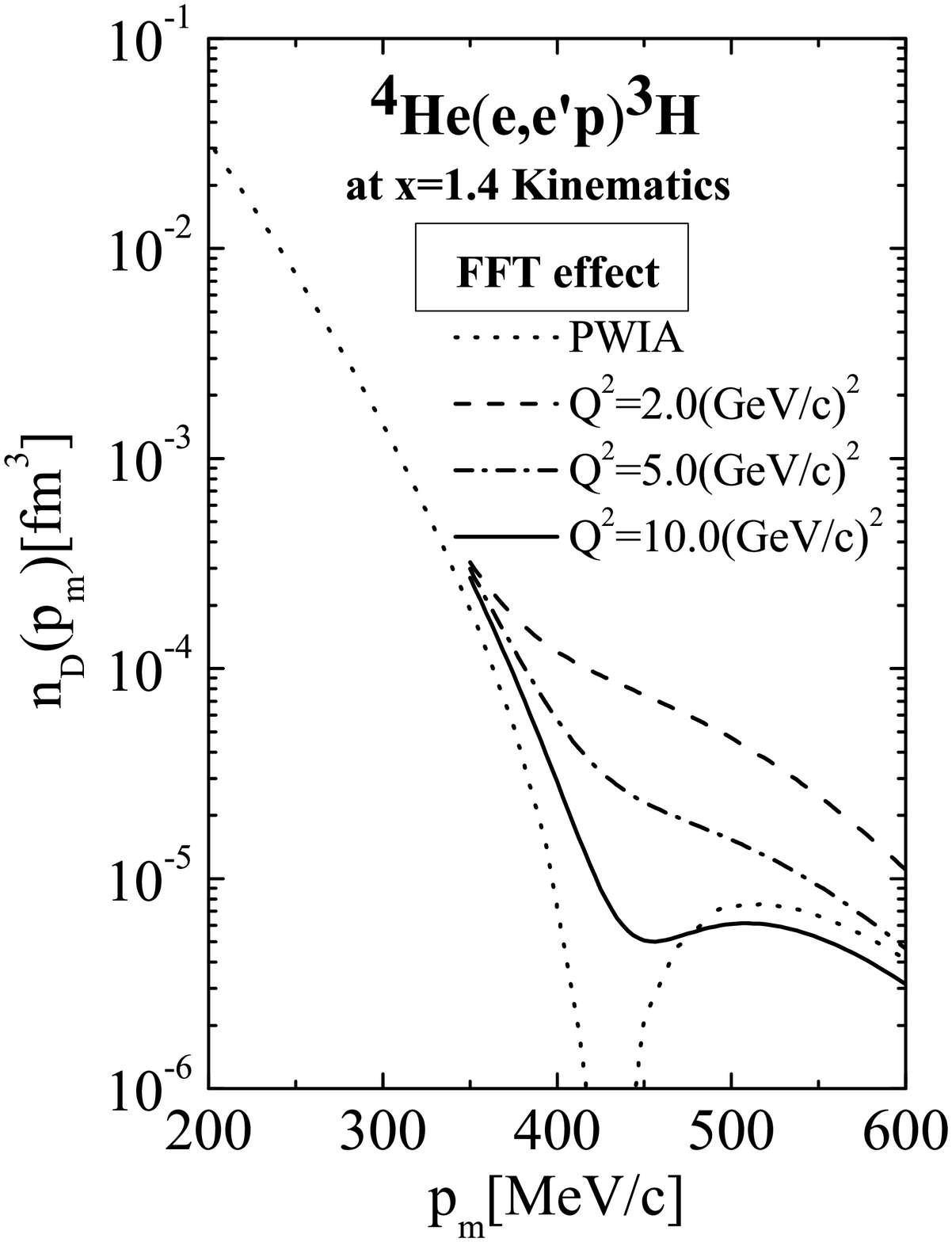}
\vskip -1.0cm \caption{The $Q^2$ dependence of FFT effects at
perpendicular kinematics with x=1.4. } \label{fig:4HeFFT2}
\end{minipage}
\end{figure}
We have calculated the cross section of the process $^4He(e,e'p)^3H$
in perpendicular kinematics introducing FFT effects (see also
\cite{FFTHiko}). The results are presented in Figs. \ref{fig:4HeFFT}
and \ref{fig:4HeFFT2}. It can be seen that at the JLAB kinematics
($Q^2=1.78\,(GeV/c)^2$, $x\sim1.8$) FFT effects, as expected,
 are too small to be detected.
We have therefore  extended  our calculation to higher values of $Q^2$
 reducing the value of $x$ to  $x$ =1.4  (in the
$CQ\omega2$ kinematics  the region with  $p_m<500\,MeV/c$ and
$Q^2\geq5\,(GeV/c)^2$ is kinematically forbidden at $x=1.8$).
  The results,  presented in
Fig. \ref{fig:4HeFFT2}, show that  FFT effects
could unambiguously be detected   in the region $5 \leq Q^2 \leq 10\, (GeV/c)^2$.
 Thus, observing the $Q^2$
dependence of the cross section of $^4He(e,e'p)^3H$ process at
$p_m\sim430\, MeV/c$ region up to around $Q^2\sim10\,(GeV/c)^2$ would be of
of great interest.
\section{Summary and Conclusions}
\label{sec:4}
 We have performed a realistic calculation of  the cross section of the
 processes  $^3He(e,e'p)^2H$, $^3He(e,e'p)(np)$ and $^4He(e,e'p)^3H$, using few-body
 wave functions which exhibit  the very
 rich correlation structure
 generated by modern NN interactions and  describing the propagation of the hit nucleon
 in the medium in term of elastic rescattering; to this end we have used
 the standard Glauber  approximation (GA),
 as  well as
   its generalized
 version (GEA). The  two approaches differ in that the latter
  takes into account in  the NN scattering amplitude the
 removal energy of the struck nucleon, or, equivalently, the excitation energy of the
 system $A-1$. Our  approach is a very  transparent one and fully  parameter free.

   The main results we have obtained,
 can be summarized as follows:
 i) the agreement between the results of our calculations and the
 experimental data for both  $^3He$
 and $^4He$, is a very satisfactory one, particularly
  in view of the lack of any adjustable parameter
 in our approach;
 ii) the effects of the FSI are such that they  systematically
  bring theoretical calculations in better agreement with the experimental data;
 for some quantities,  they simply improve the agreement
 between theory and experiment, whereas for some other quantities,
 they play a dominant role;
iii)  the 3bbu channel in $^3He$, i.e. the  process  $^3He(e,e'p)(np)$,
  provides evidence of NN correlations, in that the experimental values of
  $p_m$ and $E_m$  corresponding  to the
   maximum values of the cross section,
satisfy to a large extent the relation  predicted by the
 two-nucleon correlation mechanism  namely $E_{m}\simeq
p_m^2/4M_N + E_3$ , with the  full FSI mainly
 affecting only the magnitude of the cross section;
iv) both for $^3He$ and $^4He$ the $p_m$ dependence of the cross section exhibits
peculiar slopes which can be interpreted in terms of multiple scattering effects,
with   triple scattering in $^4He$  starting  to
significantly contribute at $p_m\geq800\,MeV/c$.
v) in the kinematical range we have considered
 only minor numerical differences were found between the conventional Glauber-eikonal
 approach and its generalized extension;
vi) finally,  we investigated the Finite Formation Time   effects, which weakens the FSI at
high $Q^2$; we found that  available data on the $^4He(e,e'p)^3H$ process are only slightly affected by
FFT effects, but, at the same time, similar data  at $Q^2 \geq 2 (GeV/c)^2$
 in the dip region ($p_m\simeq430\, MeV/c$) would provide
 a significant check of theoretical models of FFT effects.

 Final results of our calculations, including also a quantitative investigation of the limits of validity
 of the factorized cross section,  will be presented elsewhere \cite{CKM}.

 \section{Acknowledgments}
L.P.K. and H.M. are  indebted to  the University of Perugia and INFN,
Sezione di Perugia, for support and  hospitality.

\end{document}